\documentclass[abstract,amssymb,amsmath]{elbrus4}
\begin{document}

\elbrustitle{OPTICAL VIBRATIONS OF HYDROGEN IN DISORDERED PALLADIUM-GOLD ALLOYS}
\elbrusauthor{Borzov\,D.N.$^{\mathit{}}$}
\elbrusaffil{$^{\mathit{}}$\hspace{-2pt}ISSP, Chernogolovka}
\elbrusemail{tihoutrom@gmail.com}

\begin{abstract}
Direct study of the optical H vibrations of diluted Pd-H solid solutions by inelastic neutron scattering (INS) is impeded due to the very small limiting hydrogen concentration, $x_{max} \leq 0.001$ at low temperatures. 
We report study of disordered Pd$_{0.8}$Au$_{0.2}$H$_{x}$ solutions with concentrations varying from $x = 0.03$ to $0.74$. 
Our INS investigation showed that the fundamental optical H band in both concentrated and dilute Pd$_{0.8}$Au$_{0.2}$H$_{x}$ solutions consists of a sharp peak with a broad shoulder towards higher energies. 
We argue that the optical band in infinitely diluted H solutions in Pd consists of a peak with a pronounced shoulder too. 
Implications for the isotopic effects in the hydrogen solubility in Pd are considered in the present paper.
\end{abstract}

\maketitle

\noindent {\bf Introduction.} In the {\it fcc} lattice of palladium, hydrogen occupies octahedral interstitial positions with cubic symmetry. One could therefore expect that in diluted solid H solutions in Pd the fundamental band of optical H vibrations should be reduced to a narrow line of 3-fold degenerate, non-interacting isotropic local oscillators. A few attempts to test this assumption by inelastic neutron scattering (INS) gave ambiguous results due to the very small limiting hydrogen concentration, $x_{max} \leq 0.001$, of the PdH$_x$ solutions at low temperatures \cite{1,2}. 

\begin{figure}[t]
\includegraphics[width=1.00\columnwidth]{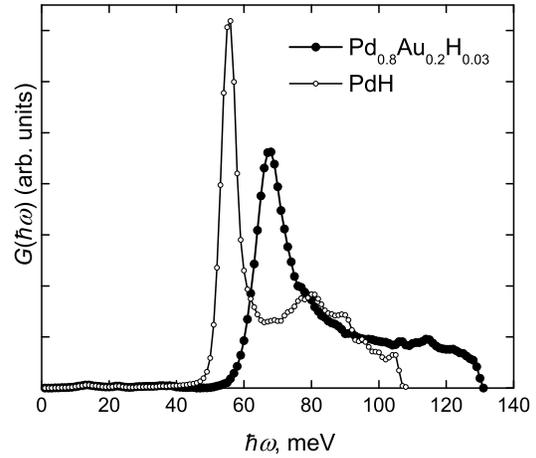}
\caption{Experimental one-phonon generalized vibrational densities of states, $G(\hbar \omega)$, corrected for the Debye-Waller factor. Open circles -- {\it fcc}~PdH measured at 25~K \cite{3}. Solid circles -- {\it fcc}~Pd$_{0.8}$Au$_{0.2}$H$_{0.03}$ measured at 5~K with the IN1-BeF spectrometer at ILL, Grenoble.}
\label{bfig1}
\end{figure}

We increased x$_{max}$ by alloying Pd with 20 at\% Au and studied disordered Pd$_{0.8}$Au$_{0.2}$H$_{x}$ solutions with concentrations varying from $x = 0.03$ to $0.74$. The samples were prepared at hydrogen pressures from 1 bar to 75 kbar. According to $^{197}$Au M\"{o}ssbauer studies \cite{3}, hydrogen atoms in the Pd$_{0.8}$Au$_{0.2}$H$_{x}$ solutions with $x \leq (0.8)^{6} \approx  0.26$ can only occupy interstices having no Au neighbours. This abates the effect of H-Au interactions in the solutions with lower H concentrations and effectively prevents hydrogen clustering at low temperatures. 
\\
\\
\noindent {\bf Inelastic neutron scattering studies.} Our INS investigation at ~5~K (IN1-BeF spectrometer, Institute Laue-Langevin, Grenoble) showed that the fundamental optical H band in both concentrated and diluted Pd$_{0.8}$Au$_{0.2}$H$_{x}$ solutions consists of a sharp peak with a broad shoulder towards higher energies. Fig.~\ref{bfig1} compares the optical spectra for the stoichiometric PdH sample \cite{3} and the Pd$_{0.8}$Au$_{0.2}$H$_{0.03}$ sample with the minimum H concentration. The shoulder amounted to approximately 0.44 integral intensity of the main peak in PdH and demonstrated no tendency to vanish with decreasing H concentration in the Pd$_{0.8}$Au$_{0.2}$H$_{x}$ samples.

Both gold and hydrogen expand the lattice of palladium. If the position $\hbar \omega_0$ of the main optical peak of the PdH and Pd$_{0.8}$Au$_{0.2}$H$_{x}$ solutions is plotted as a function of the atomic volume, its linear extrapolation to the volume of pure Pd metal gives $\hbar \omega_0 = 69.1$ meV. This well agrees with an experimental value of $68.5 \pm 0.5$ meV for PdH$_{0.002}$ \cite{1} and therefore shows that the extrapolation procedure is meaningful. The extrapolation also shows that the spectrum Pd$_{0.8}$Au$_{0.2}$H$_{0.03}$ should well represent the fundamental optical band of the infinitely diluted Pd(H) solution. The position of the centre of gravity of this band gives the weight-averaged energy $\hbar \omega_{\textrm{aver}} \approx 80$~meV for optical vibrations in Pd(H).

One of the possible explanations of the high-energy shoulder in the spectra of diluted hydrogen solutions in palladium is the smearing of the peak due to what can be classically interpreted as the changes in the high-frequency vibrations of the H atom in a slowly varying ``instant'' potential of the slowly moving heavy metal atoms. Mention in this connection that a molecular-dynamics calculation for an H atom in the Pd matrix \cite{4} gave a Fourier transform of the time correlation function that qualitatively reproduces the experimental peak with a broad shoulder in the INS spectra of the diluted Pd-Au-H solutions.
\\
\begin{figure}[t]
\includegraphics[width=1.0\columnwidth]{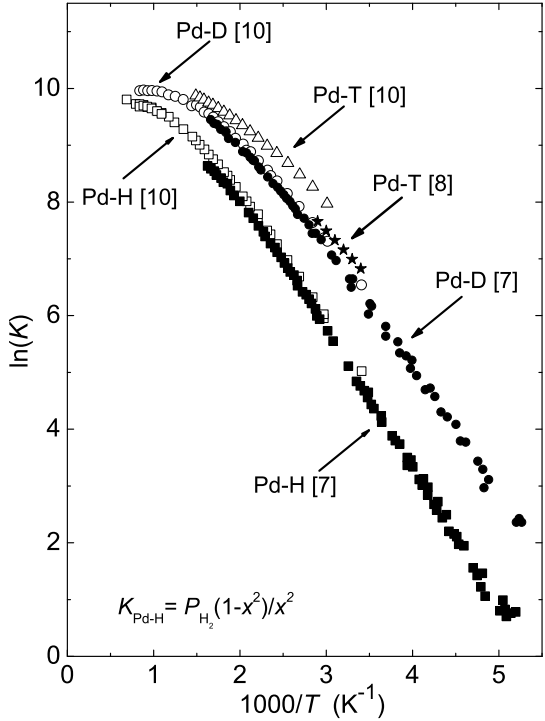}
\caption{Available experimental data on the solubility of hydrogen isotopes in Pd in the limit of infinite dilution.}
\label{bfig2}
\end{figure}

\noindent {\bf Implications for the isotopic effects in the hydrogen solubility in palladium.} The weight-averaged energy $\hbar \omega_{\textrm{aver}}$ of optical hydrogen vibrations plays an important role in the thermodynamics of the Pd-H system. The value $\hbar \omega_0 = 68.5$~meV of the main peak \cite{1} mistakenly used in the calculations instead of  $\hbar \omega_{\textrm{aver}} \approx 80$~meV since 1976 has caused much difficulty in the interpretation of experimental results and, in particular, led to the ill-founded conclusion that the optical hydrogen vibrations in Pd are strongly anharmonic. 

Assuming that the spectrum of Pd$_{0.8}$Au$_{0.2}$H$_{0.03}$ depicted in Fig.~\ref{bfig1} represents the phonon density of states, $g(\omega)$, of the infinitely diluted Pd(H) solution, we reanalyzed the experimental results in literature concerning the protium, deuterium and tritium solubility in palladium and demonstrated that these results agree with the harmonic model.

The literature results are shown in Fig.~\ref{bfig2} in the form of temperature dependences of the Sieverts constant determined as $K_{\rm H} =\left[ {\rm H}_2 \right] / \left[ {\rm H} \right]^2 = \mbox{$P_{\rm H_2}(1-x)^2 / x^2$}$ for the investigated diluted PdH$_x$ solutions and in a similar way for the PdD$_x$ and PdT$_x$ solutions. In the case of infinitely diluted solutions, one can write:
$$
\ln{K_{\rm H}(T)}=(2G_{\rm H}^{\infty}(T) - G_{{\rm H}_2}(T,P_0))/RT,
$$
\noindent where $G_{{\rm H}_2}(T,P_0)$ is the standard Gibbs energy of H$_2$ gas and $G_{\rm H}^{\infty}(T)$ is the Gibbs energy of 1 mol of H atoms in Pd at infinite dilution. Subtracting this equation from an analogous equation for the Pd-D solutions yields the expression:
\begin{multline}
\Delta G_{\rm DH} / RT = \left[G_{\rm D}^{\infty}(T)-G_{\rm H}^{\infty}(T) \right] / RT =  \\
\frac{1}{2} \ln \left [{K_{\rm D} (T)/K_{\rm H} (T)}\right ] + \frac{1}{2} \left[ G_{\rm D_2}-G_{\rm H_2} \right] / RT 
\label{eq1}
\end{multline}

\begin{figure}[t]
\includegraphics[width=1.00\columnwidth]{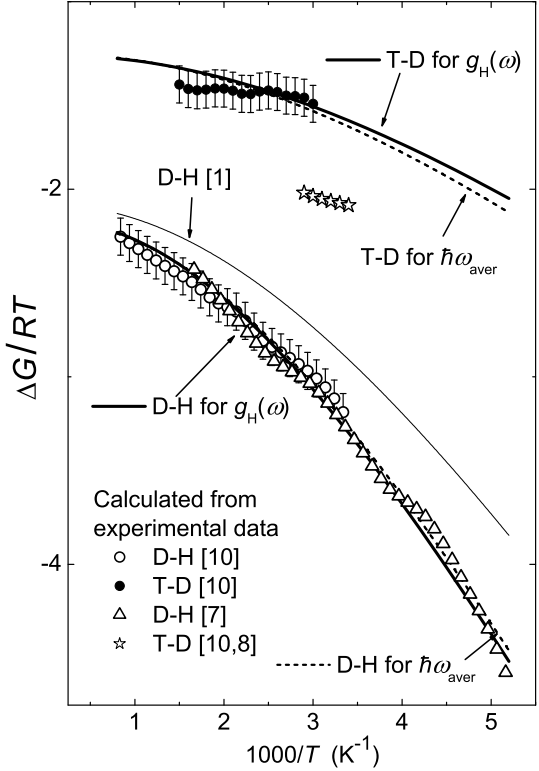}
\caption{The dimensionless differences $\Delta G_{\rm DH} / RT$ and $\Delta G_{\rm TD} / RT$ calculated using the experimental data from Fig.~\ref{bfig2} and results of fitting these differences by modelling the dynamics of hydrogen isotopes in Pd metal (solid and dashed lines, see text). }
\label{bfig3}
\end{figure}

The standard Gibbs energy for the gaseous phase of each hydrogen isotope has been accurately determined and tabulated and is available in the literature \cite{6}. It is therefore advisable to calculate the right-hand side of equation (\ref{eq1}) and to consider it as the experimental $\Delta G_{\rm DH}^{exp} / RT$ dependence to be fitted by the difference $\left[G_{\rm D}^{\infty}(T)-G_{\rm H}^{\infty}(T) \right] / RT$ that only depends on the properties of the Pd(D) and Pd(H) solutions. The temperature dependences of $\Delta G_{\rm DH}^{exp} / RT$ and also $\Delta G_{\rm TD}^{exp} / RT$ are presented in Fig.~\ref{bfig3} by symbols.

Due to the large ratio of the atomic masses of hydrogen and palladium $(m_{\rm Pd} = 106.4$~u), the difference $G_{\rm D}^{\infty}(T)-G_{\rm H}^{\infty}(T)$ is nearly solely determined by the difference in the free energies of optical vibrations of D and H atoms, the energies being counted from the bottom of the potential well for these atoms in the Pd(D) and Pd(H) solutions, correspondingly \cite{9}. H and D atoms in Pd metal were earlier considered as three-dimensional Einstein oscillators because the main optical peaks had only been observed in the INS spectra of diluted PdH$_x$ and PdD$_x$ solutions \cite{1}. In the Einstein model:
\begin{multline}
\left[ G_{\rm D}^\infty (T)-G_{\rm H}^\infty (T) \right] / RT \approx \\ 3 \ln \left[ \left( 1-\exp \left(-\frac {\hbar \omega_{\rm D}}{kT}\right )\right )\biggm /\left( 1-\exp \left(-\frac {\hbar \omega_{\rm H}}{kT}\right )\right ) \right] \\ + \frac{3}{2} (\hbar \omega_{\rm D} - \hbar \omega_{\rm H}) / kT + \Delta_{\rm DH} /kT
\label{eq2}
\end{multline}
\noindent where $\hbar \omega$ is the energy of the first harmonics, $k$ is the Boltzmann constant and $\Delta_{\rm DH}$ is the difference in the bottom energies of the potential wells for D and H atoms. The usage of $\hbar \omega_{\rm H} = 68.5$~meV and $\hbar \omega_{\rm D} = 48$~meV from ref.~\cite{1} gives the dependence shown in Fig.~\ref{bfig2} by the thin solid line labeled as ``D-H [1]''. The line irrecoverably disagrees with experiment. 

If the densities of optical phonon states, $g_{\rm H} (\omega)$ and $g_{\rm D} (\omega)$, are known for the infinitely diluted Pd(H) and Pd(D) solutions, one can write 
\begin{multline}
\left[G_D^\infty (T)-G_{\rm H}^\infty (T) \right] / RT \approx \\ 
3 \int\nolimits_0^\infty \ln[1-\exp (- \hbar \omega / kT)] [g_{\rm D} (\omega) - g_{\rm H} (\omega)] d \omega + \\
\frac{3}{2} \int\nolimits_0^\infty \hbar \omega  [g_{\rm D} (\omega) - g_{\rm H} (\omega)] d \omega /kT +  \Delta_{\rm DH} /kT,
\label{eq3}
\end{multline}
\noindent where the $g (\omega)$ spectra are normalized so that $\int g (\omega) d \omega =1$.

Assuming that $g_{\rm H} (\omega)$ is proportional to the generalized density of states for the Pd$_{0.8}$Au$_{0.2}$H$_{0.03}$ sample shown in Fig.~\ref{bfig1} and that $g_{\rm D} (\omega) = r g_{\rm H} (r \omega)$, we fitted $\Delta G_{\rm DH}^{exp} / RT$ with equation (\ref{eq3}) using $r$ and $\Delta_{\rm DH}$ as fitting parameters. The resulting dependence is shown in Fig.~\ref{bfig3} by the thick solid line labeled "D-H for $g(\omega)$". As one can see, it well agrees with experiment. The obtained optimum value of $r = \sqrt{2.06} \approx \sqrt{m_{\rm D} / m_{\rm H}}$ suggests the nearly harmonic character of optical vibrations in Pd(H) and Pd(D). The optimum value of $\Delta_{\rm DH} = -3.3$~meV indicates that the bottom of the potential well for H atoms in Pd lies deeper than for D atoms by 3.3~meV. 

Fitting the $\Delta G_{\rm TD}^{\textrm{exp}} / RT$ from ref.~\cite{7} in a similar way yields a rather good approximation to experiment (upper thick solid line in Fig.~\ref{bfig3}) with $r = \sqrt{1.52} \approx \sqrt{m_{\rm T} / m_{\rm D}}$ and $\Delta_{\rm TD} \approx 0$. The $\Delta G_{\rm TD}^{\textrm{exp}} / RT$ dependence from ref.~\cite{8} (open stars in Fig.~\ref{bfig3}) is less accurate than that from ref.~\cite{7} and it cannot be fitted with physically acceptable values of the fitting parameters. 

Interestingly, the $\Delta G_{\rm DH}^{\textrm{exp}} / RT$ and $\Delta G_{\rm TD}^{\textrm{exp}} / RT$ dependences can be well approximated with equation (\ref{eq2}) (see dashed lines in Fig.~\ref{bfig3}) using virtually the same values of $r$ and $\Delta$ as above, if $\hbar \omega_{\rm H}$ is set equal to the weight-averaged energy $\hbar \omega_{\textrm{aver}} = 80$~meV of optical vibrations in Pd(H). 
\\

\noindent {\bf Conclusions.} The inelastic neutron scattering investigation of the Pd-Au-H solutions has for the first time given the opportunity to establish the spectrum of optical vibrations in diluted Pd-H solutions.  

In contrast to the commonly held view, the band of optical H vibrations in the Pd-H solutions does not shrink to a narrow line of 3-fold degenerate local oscillators on infinite dilution. Instead, the fundamental optical peak has a broad and intense shoulder towards higher energies. As a result, the weight-averaged energy $\hbar \omega_{\textrm{aver}}$ of optical vibrations in diluted Pd-H solutions is equal to about 80~meV and not to 68.5~meV as it was assumed earlier.

The value of $\hbar \omega_{\textrm{aver}} = 80$~meV well describes all available experimental data on the H and D solubility in Pd in the harmonic approximation. This affords no basis for the widely spread opinion that optical vibrations in the diluted Pd-H solutions are strongly anharmonic.   

The analysis of the solubility of hydrogen isotopes in palladium shows that the bottom of the potential well is located at approximately the same energy for the D and T atoms and lies by some 3.3~meV deeper for the H atoms. The force constants are a little higher for the H atoms than for the D atoms and coincide for the D and T atoms within the experimental error.
\\
\\
\noindent {\bf Acknowledgements.} This work was supported by grant No. 08-02-00846 from the Russian Foundation for Basic Research and by the Program ``Physics of Strongly Compressed Matter'' of the Russian Academy of Sciences.

\end{document}